\documentstyle[times,emulateapj]{article}
\newcommand{\PSbox}[3]{\mbox{\rule{0in}{#3}\includegraphics{#1}\hspace{#2}}}

\begin{document}
 
\title{{\it RXTE} All-Sky Monitor Localization of SGR 1627--41}

\author{Donald A. Smith\altaffilmark{1}, Hale V. Bradt\altaffilmark{1}, and Alan M. Levine\altaffilmark{1}}

\altaffiltext{1}{Center for Space Research and Department of Physics, MIT, Cambridge, MA 02139}

\authoremail{dasmith@space.mit.edu}

\begin{abstract}

The fourth unambiguously identified Soft Gamma Repeater (SGR),
SGR~1627--41, was discovered with the BATSE instrument on 1998 June 15
(\markcite{kouve98}Kouveliotou et al. 1998).  Interplanetary Network
(IPN) measurements and BATSE data constrained the location of this new
SGR to a $6\arcdeg$ segment of a narrow ($19\arcsec$) annulus
(\markcite{hurl99a}Hurley et al. 1999a; \markcite{woods98}Woods et
al. 1998).  We present two bursts from this source observed by the
All-Sky Monitor (ASM) on {\it RXTE}.  We use the ASM data to further
constrain the source location to a $5\arcmin$ long segment of the
BATSE/IPN error box.  The ASM/IPN error box lies within $0.3\arcmin$
of the supernova remnant (SNR) G337.0--0.1.  The probability that a
SNR would fall so close to the error box purely by chance is
$\sim5$\%.

\end{abstract}
 
 
\section{Introduction}

The soft gamma repeaters (SGRs) were first identified as a separate
class from ``classical'' gamma-ray bursts over ten years ago
(\markcite{attei87}Atteia et al. 1987; \markcite{kouve87}Kouveliotou
et al. 1987; \markcite{laros87}Laros et al. 1987).  To date, four SGRs
have been unambiguously identified.  Attempts have been made to
associate all these sources with supernova remnants (SNR).  The
location of SGR~0525--66 is consistent with the SNR~N49
(\markcite{cline82}Cline et al. 1982), and SGR~1806--20 has been
associated with the SNR G10.0--0.3 (\markcite{kulka93}Kulkarni \&
Frail 1993; \markcite{kouve94}Kouveliotou et al. 1994;
\markcite{kulka94}Kulkarni et al. 1994; \markcite{murak94}Murakami et
al. 1994).  SGR~1900+14 has recently been localized to an error box of
1.6 square arcminutes, which lies just outside SNR G42.6+06
(\markcite{hurl99b}Hurley et al. 1999b).

SGR~1627--41 was discovered with the BATSE instrument on 1998 June~15;
a coarse location was promptly announced
(\markcite{kouve98}Kouveliotou et al. 1998).  Three days later, Hurley
\& Kouveliotou \markcite{hurle98}(1998) reported, based on data from
GRB detectors in the Interplanetary Network (IPN), that the burst
source was located within an annulus $6\arcmin$ in width.  Further
analysis reduced the width of the annulus to $19\arcsec$
(\markcite{hurl99a}Hurley et al. 1999a).  Earth limb considerations
during bursts observed with BATSE restricted the burst source location
along the IPN annulus to declinations between $-43\arcdeg$ and
$-49\arcdeg$ (\markcite{woods98}Woods et al. 1998).  These
localizations are all displayed in Figure~\ref{disc}.  Woods et
al. \markcite{woods98}(1998) noted that the non-thermal core of the
CTB~33 complex lay within this region.  This core was identified as
SNR G337.0--0.1 by Sarma et al. \markcite{sarma97}(1997), who
estimated its distance to be $11.0\pm0.3$ kpc.  In response to the
discovery of SGR~1627--41, this SNR was observed with the {\it
BeppoSAX} Narrow Field Instruments.  In this observation, a faint
X-ray source was discovered at a location consistent with the IPN
annulus, on the west side of G337.0--0.1 (\markcite{woods99}Woods et
al. 1999).

In this paper, we use observations by the {\it Rossi X-ray Timing
Explorer} All-Sky Monitor of two bursts from SGR~1627-41 to constrain
the source's position along the IPN annulus, and we discuss the
possible association of SGR~1627--41 with the SNR G337.0--0.1.

\setcounter{figure}{0} 
\refstepcounter{figure} 
\PSbox{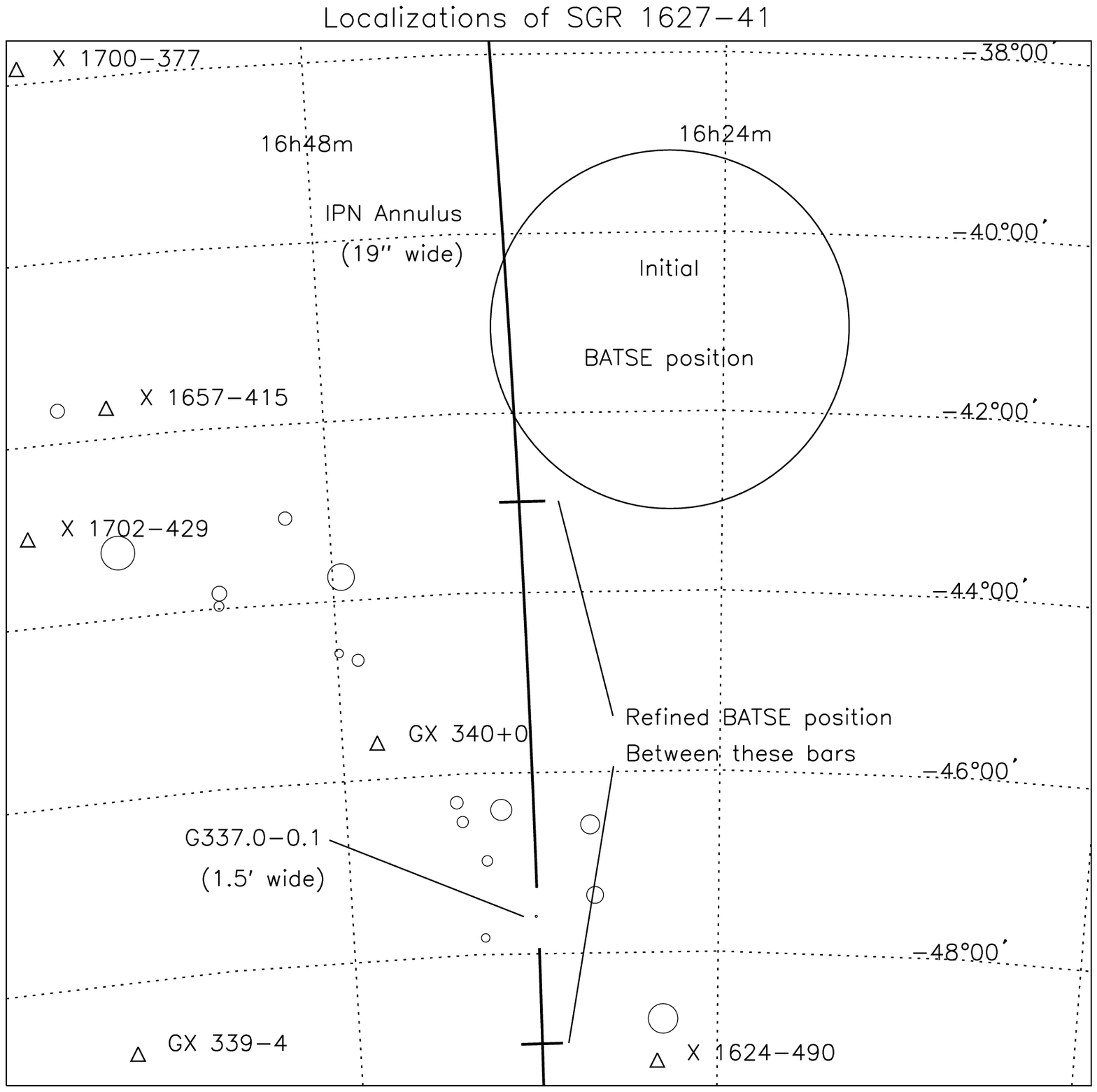 hoffset=-15 voffset=0 hscale=52
vscale=52}{8.8cm}{9.5cm}{\\\\\small Fig. 1 -- Localizations of
SGR~1627--41.  Known X-ray sources in the ASM catalog with typical
brightnesses above $\sim50$~mCrab are plotted as triangles.  SNRs from
the Green catalog are plotted as circles with diameters equal to the
mean angular size listed in the catalog.  The IPN annulus is not
plotted near the SNR G337.0--0.1 to avoid obscuring the $1.5\arcmin$
circle.\label{disc}}

\section{Instrument}

The ASM consists of three Scanning Shadow Cameras (SSCs) mounted on a
motorized rotation drive.  Each SSC contains a proportional counter
with eight resistive anodes that are used to obtain a one-dimensional
position along the direction parallel to the anodes for each detected
event.  The proportional counter views a $12\arcdeg \times 110\arcdeg$
(FWZI) field through a random-slit mask.  The mask consists of six
unique patterns, oriented such that the slits run perpendicular to the
anodes.

The intensities of known sources in the field of view (FOV) are
derived via a fit of model slit-mask shadow patterns to histograms of
counts as a function of position in the detector.  Normally, the
residuals from a successful fit in the 1.5--12 keV band are then
cross-correlated with each of the expected shadow patterns
corresponding to a set of possible source directions which make up a
grid covering the FOV.  A peak in the resulting cross-correlation map
indicates the possible presence and approximate location of a new,
uncataloged X-ray source (\markcite{levin96}Levine et al. 1996).

In addition to ``position'' data products, time-series data on the
total number of counts registered in each SSC are recorded in 1/8~s
bins.  The position of each count in the detector is not preserved in
the time-series mode.

\section{Observations and Analysis}

Two bursts were detected in SSC~3 of the ASM on 1998 June 17.943917
and 17.954243 (UTC).  The time histories of these bursts are shown in
Figure~\ref{mtsfig}.  Each burst was short ($\lesssim2$ s), bright
(5--12~keV fluxes of $\sim9$ and $\sim23$~Crab at peak), and hard (no
significant flux below 5~keV), characteristic of SGR bursts.  At these
times, BATSE could not observe SGR~1627--41 due to Earth-occultation
(C. Kouveliotou 1998, private communication), but SGR~1627--41 was
known to be active around this time, and the BATSE and IPN
localizations of SGR~1627--41 were consistent with the FOV of the ASM
at the times of these events.  We therefore attribute these bursts to
SGR~1627--41.

The energy spectra of bursts from SGR~1806--20 are known to drop
rapidly below $\sim14$~keV (\markcite{fenim94}Fenimore, Laros, \&
Ulmer 1994), and this property seems consistent with the ASM
observations of SGR~1627--41 reported here.  These two bursts were not
detected in the lowest two energy channels of the ASM (1.5--5~keV).
We therefore performed the cross-correlation analysis using only the
data from the highest energy channel (5--12~keV).  We found
significant peaks in each of the two cross-correlation maps.  The
celestial locations of these peaks are consistent with each other and
with the refined BATSE/IPN position.  The first burst detection has a
statistical significance of 4.3~$\sigma$, while the second burst was
detected with a significance of 5.7~$\sigma$.  These levels of
significance do not take the number of trials in the search into
account.

Since there are roughly 10000 independent position bins in the FOV of
an SSC, the probability of measuring a noise peak of at least
4.3~$\sigma$ somewhere in the FOV is roughly 0.08, and the probability
of measuring a 5.7~$\sigma$ peak or higher is roughly $10^{-3}$.
There are roughly 60 independent position bins within the refined
BATSE/IPN error box.  The probability that both peaks would fall at
random in the same location and that the common location should
overlap the BATSE/IPN error box is $60 \times (10^{-4})^2 \sim
10^{-6}$.  We are therefore confident that these peaks do represent
detections of SGR~1627--41.

To refine the source position, we determined which regions of the IPN
annulus were consistent with the ASM observations of the two bursts.
At the time of the first burst, the FOV of SSC~3 did not cover
positions along the annulus north of $-45\arcdeg$, while at the time
of the second burst, the FOV did not cover positions along the annulus
south of $-50.6\arcdeg$.  We ran multiple trials of our ASM
shadow-pattern fitting program for the 5--12~keV data from each of the
two 90~s observations.  In each trial, a source was assumed to be at
one of 1600 locations along the center line of the segment of the IPN
annulus with $-50\arcdeg \leq \delta \leq -45\arcdeg$.  

\refstepcounter{figure} 
\PSbox{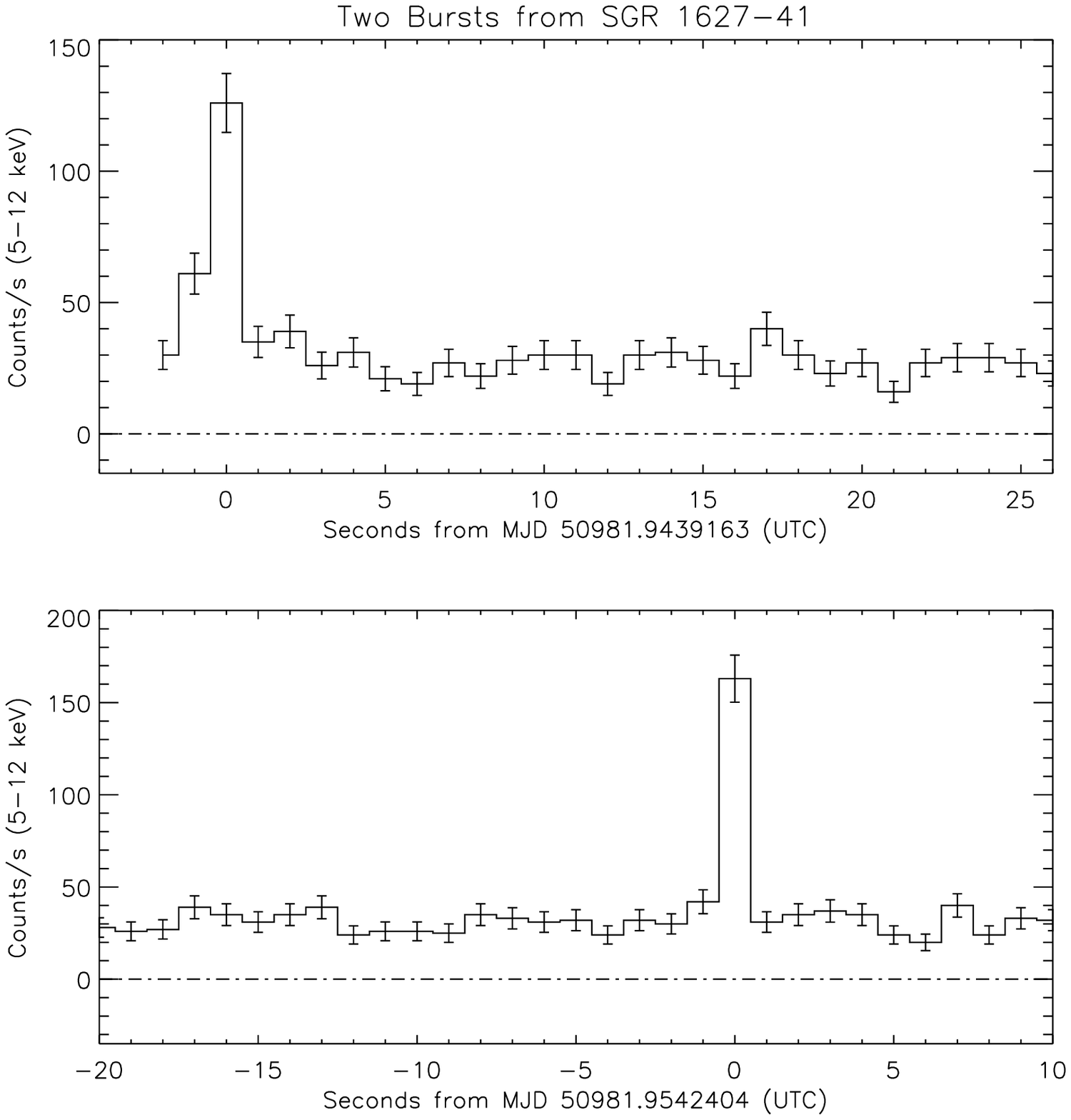 hoffset=-25 voffset=0 hscale=54
vscale=54}{8.8cm}{9.7cm}{\\\\\small Fig. 2 -- Counts per 1-s bin
around the times of two bursts from SGR~1627--41, as observed between
5--12~keV by SSC~3.  The count rate includes contributions from all
X-ray sources in the FOV, as well as the diffuse X-ray background.  No
background subtraction has been performed.\label{mtsfig}}

\vspace{0.4cm}

\noindent
Each trial yields the fitted intensity of the source as well as a
$\chi^2$ goodness of fit statistic.  For both observations, the best
fit was found to be near $\delta = -47.6\arcdeg$ (Fig.~\ref{dipfig}).

For each of the two ASM observations, regions for which $\chi^2 <
\chi^2_{\rm min} + $ 2.7, 4.0, or 6.6 yielded 90\%, 95\%, or 99\%
confidence intervals, respectively, for the source declination.  These
intervals are presumed to reflect counting statistics.  We estimated
the effect of systematic errors by projecting the measured magnitude
of the position error for strong sources onto the direction of the
BATSE/IPN error box.  This magnitude was measured by Smith et
al. \markcite{smith99}(1999) to be $1.9\arcmin$ for 95\% confidence
intervals along the direction parallel to the proportional counter
anodes.  We added $1.9\arcmin/\cos{\theta}$ in quadrature with the
errors estimated from the $\chi^2$ values, where $\theta$, the angle
between the IPN annulus and the anode direction, was $7\arcdeg$ for
the first burst and $43\arcdeg$ for the second burst.

We thus derive two independent measurements of the declination of
SGR~1627--41 along the IPN annulus: $\delta_1 =
-47\arcdeg.603^{+0.053}_{-0.051}$ and $\delta_2 =
-47\arcdeg.640^{+0.091}_{-0.096}$ at 95\% confidence.  We find a joint
error box of $<\delta> = -47\arcdeg.621\pm0.045$ (J2000) by taking the
weighted average of the two measurements.  We calculate similar
intervals for 90\% and 99\% confidence levels, using the same value
for the systematic error.  These three intervals are $4.9\arcmin$,
$5.4\arcmin$, and $6.3\arcmin$ long, in order of increasing
confidence.  All three intervals are plotted as dark bars in
Figure~\ref{dipfig} and, together with the IPN annulus, as boxes in
Figure~\ref{radfig}.  The center of these boxes lies $0.13\arcdeg$
below the galactic plane.  These boxes are consistent with the
$1\arcmin$ location of the persistent X-ray source localized by Woods
et al. \markcite{woods99}(1999), which is also plotted in
Figure~\ref{radfig}.

\refstepcounter{figure} 
\PSbox{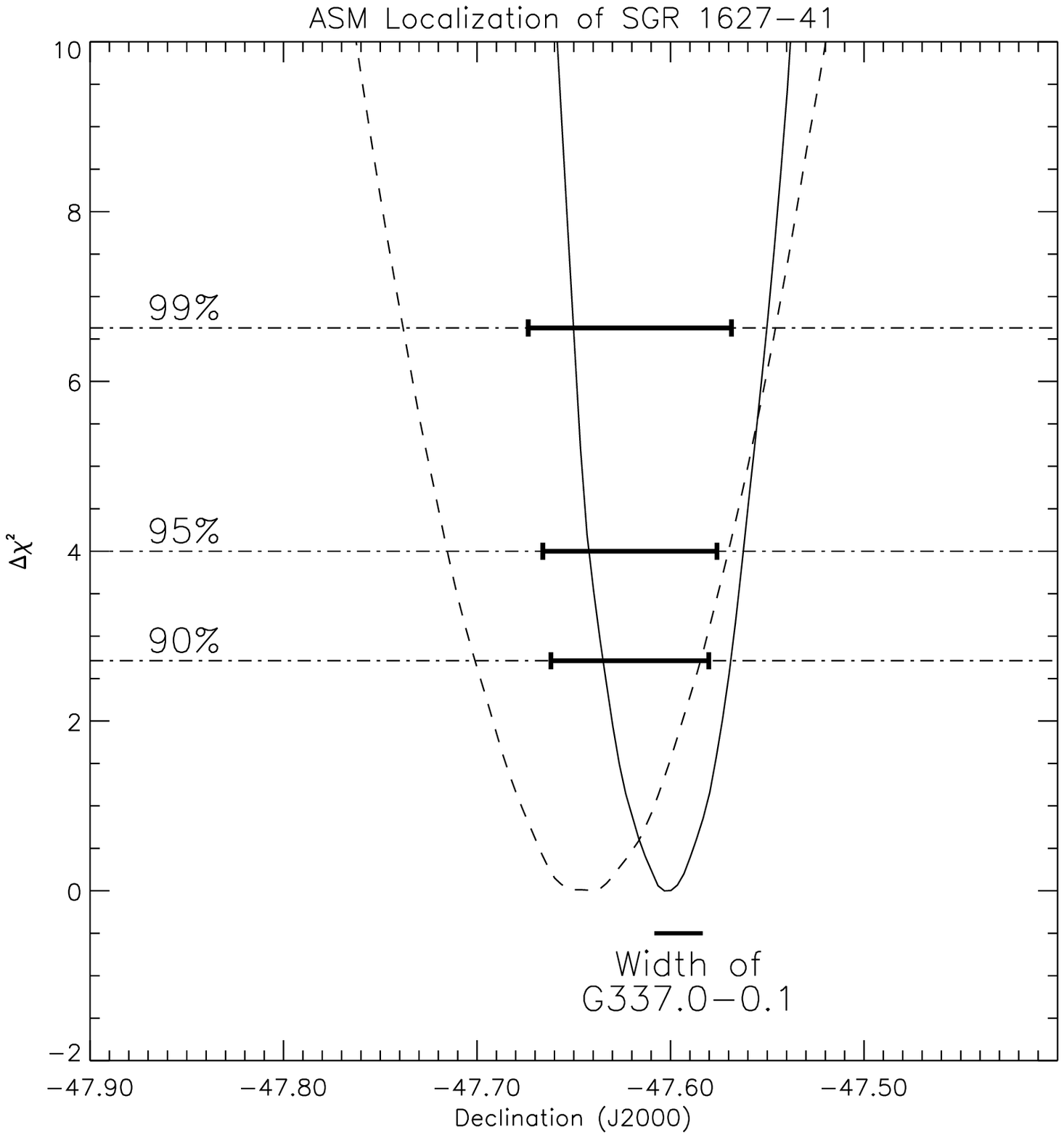 hoffset=-25 voffset=-7 hscale=55
vscale=55}{8.8cm}{9.7cm}{\\\\\small Fig. 3 -- The change in $\chi^2$
as a function of declination along the IPN annulus, relative to the
minimum values.  The results from the burst of June~17.943917 are
graphed as a solid line, while those from June~17.954243 are graphed
as a dashed line.  The location and extent of G337.0--0.1 is indicated
by the dark horizontal bar below the curves.  The weighted average
error box at each confidence level is indicated by a darkened
interval.\label{dipfig}}

\vspace{0.4cm}

Since the position of the source is well-constrained, the fits to the
position data yield source intensities that can be used to estimate
the number of counts actually detected from the source.  Since there
are no other known variable sources in the FOV, we can use the
time-series data to estimate the peak flux of the bursts as well as
investigate the possible presence of non-burst emission. Comparison of
the count rates with the observed brightness of the Crab Nebula yields
burst peak fluxes (1-s~bins; 5--12~keV) of $(1.1\pm0.2)\times10^{-7}$
and $(2.7\pm0.3)\times10^{-7}$~erg~cm$^{-2}$~s$^{-1}$.  These values
may be underestimates because the burst spectrum is substantially
harder than the Crab spectrum in this energy range.  We find that, for
both bursts, the number of counts derived from the position data is
consistent with the number of counts detected in the bursts as
recorded in the time-series data.  This yields a weak upper limit of
$\sim2\times10^{-8}$~erg~cm$^{-2}$~s$^{-1}$ (3~$\sigma$) on any
non-burst emission in the 1.5--12~keV band from the SGR during these
observations.

A search for pulsations in the 5--12~keV time-series data was
conducted by performing FFTs on 64~s of data after the burst events.
In neither observation was any coherent signal between 0.015 and
4.000~Hz detected to an upper limit on the amplitude of $\sim2.4$~c/s
at 95\% confidence.  At the position of the ASM/IPN error box, this
limit corresponds to a peak-to-peak modulation of $\sim340$~mCrab.

\section{Discussion}

The $19\arcsec \times 4.9\arcmin$ ASM/IPN error box for the bursting
source passes within $1.05\arcmin$ of the center of SNR G337.0--0.1,
which is 

\refstepcounter{figure} 
\PSbox{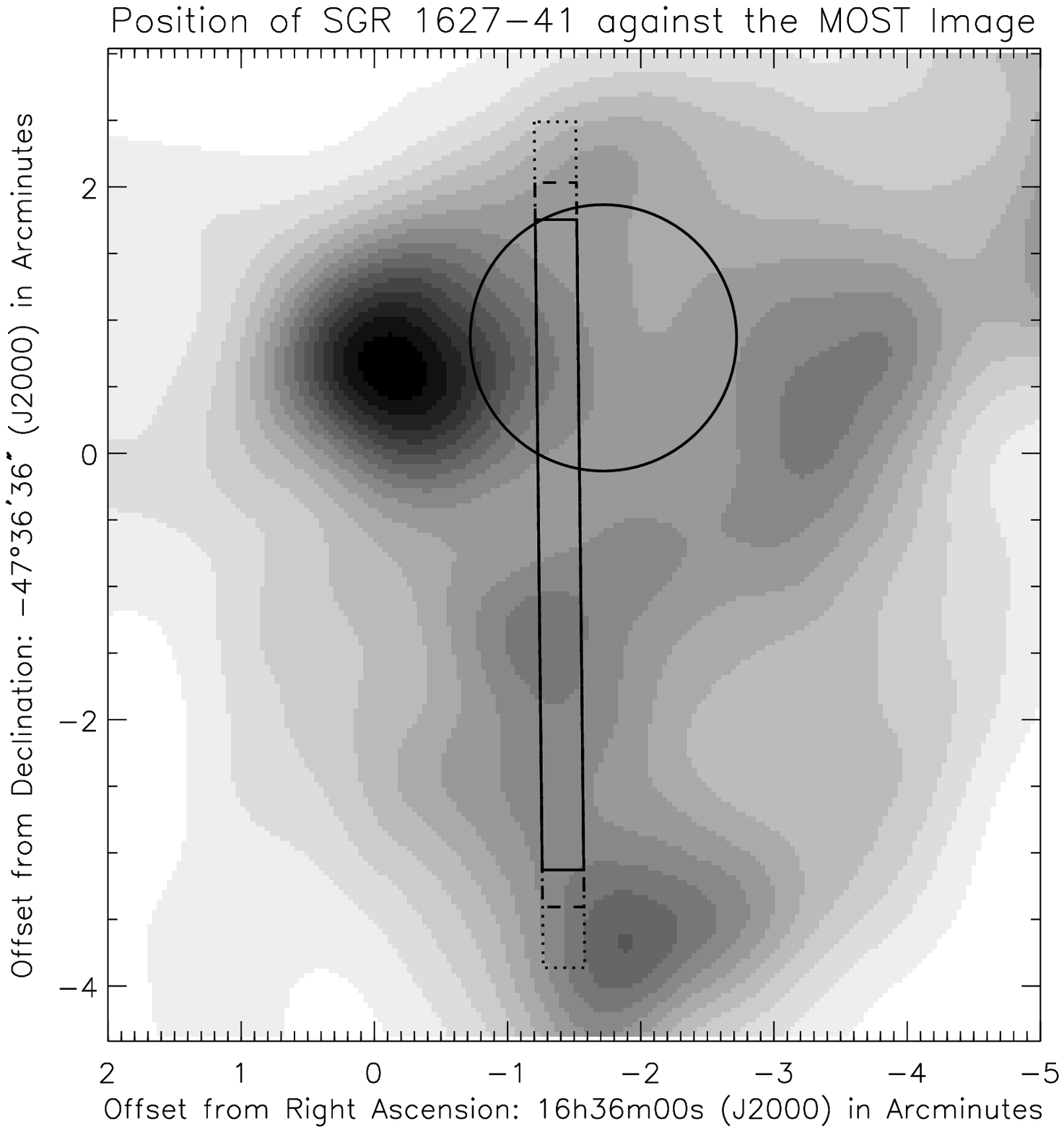 hoffset=-25 voffset=-7 hscale=54
vscale=54}{8.8cm}{9.7cm}{\\\\\small Fig. 4 -- The joint ASM/IPN
localization of the bursts from SGR~1627--41 is graphed over the MOST
0.8~GHz image (Whiteoak \& Green, 1996) of SNR G337.0--0.1, the
pronounced dark region.  The width of the box is set by the IPN
3~$\sigma$ error annulus.  The lengths represent three confidence
levels for the ASM localization: 99\% (dotted line), 95\% (dashed
line), and 90\% (solid line).  Also plotted is the error circle for
the location of the weak persistent source observed by {\it BeppoSAX}.
(Woods et al. 1999).\label{radfig}}

\vspace{0.4cm}

\noindent
a non-thermal shell, $0.75\arcmin$ in mean radius, embedded in a
complex HII region (\markcite{sarma97}Sarma et al. 1997).  Since
arguments have been made to associate each of the three previously
known SGRs with nearby SNRs, it is tempting to conclude that
SGR~1627--41 is a magnetar that was born in the same supernova
explosion that produced SNR G377.0--0.1.  However, this is a very
crowded region of the sky.  The {\it a posteriori} probability of
finding a SNR near the error box for SGR~1627--41 cannot be
discounted, and it has been argued (e.g., \markcite{gaens95}Gaensler
\& Johnston 1995) that the number of pulsar/SNR associations has been
overestimated due to the underestimation of the chances of appearing
close on the sky by geometric projection.

The north end of the ASM/IPN error box lies $0.3\arcmin$ outside the
nominal edge of G337.0--0.1.  We estimate the probability of a chance
association after the method of Kulkarni \& Frail
\markcite{kulka93}(1993).  We inflate the size of all the SNRs in
Green's catalog \markcite{green98}(1998) by $0.3\arcmin$ and convolve
them with the ASM/IPN error box for SGR~1627--41.  The fraction of the
sky covered by the total of all the resulting areas gives the
probability of a chance association, if the distribution of SNRs is
uniform.  The ASM/IPN error box is less than $0.1\arcdeg$ away from
the galactic plane, and SNRs are strongly clustered along the galactic
plane as well as toward the galactic center.  We therefore apply the
method by summing over the areas of the 19 SNRs in the region with
galactic coordinates of $327\arcdeg < \ell < 347\arcdeg$ and $\mid b
\mid < 0.5$.  We obtain a probability of 5.4\% that the ASM/IPN error
box would fall within $0.3\arcmin$ of a SNR by chance.

Another method of evaluating the association between SGR~1627--41 and
G337.0--0.1 is to consider the conditional probability of finding the
SGR within the ASM error box given that it must be somewhere within
the revised BATSE/IPN error box.  The hypothesis that the SGR was born
in the same explosion that created G337.0--0.1 gives us one model for
the underlying probability distribution of the source location.  We
assume that the SGR has been traveling for $10^4$~y from the center of
G337.0--0.1 (assumed to be 11~kpc from Earth) at a velocity drawn from
a three-dimensional Maxwellian distribution with an rms value of
500~km/s.  We convert this three-dimensional distribution to a
distribution on the sky (a two-dimensional Gaussian with a standard
deviation of $1.6\arcmin$), and renormalize this function by requiring
that its integral over the BATSE/IPN error box be equal to unity.  The
integral of the resulting density function over the ASM 90\%
confidence error box yields a value of~0.749.

To evaluate a second model, we compute the probability of finding
the SGR within the ASM error box if its location within the BATSE/IPN
error box is drawn from a uniform distribution.  Under that
assumption, the probability is simply the ratio of the areas of the
two error boxes, or~0.0136.

A comparison between these two probabilities indicates that the first
hypothesis is a more reasonable explanation for the data than the
second.  This should not be taken as proof that the SGR actually did
originate at the center of SNR G337.0--0.1, as it is also possible
that the SGR may be associated with something other than G337.0--0.1
that belongs to a class of objects that clusters along the galactic
plane.

An association between SGR~1627--41 and G337.0--0.1 does not require
that unreasonable physical characteristics be attributed to the
system.  At an assumed distance of 11~kpc, the ASM/IPN error box lies
4~pc away from the center of G337.0--0.1, projected onto the plane of
the sky.  G337.0--0.1 itself is 5.1~pc in diameter.  The radius of a
SNR in the Sedov-Taylor phase of its expansion is given by $R = (31.5\
{\rm pc})(E_{51}/n_0)^{1/5}t_5^{2/5}$ (\markcite{sfs89}Shull, Fesen,
\& Saken 1989).  If we assume $E_{51}\sim1$, then $t_5 =
n_0^{1/2}(R/31.5)^{5/2}$.  Since OH masers have been associated with
this SNR (\markcite{frail96}Frail et al. 1996), the local density must
fall within the range $(1-30)\times10^4$~cm$^{-3}$
(\markcite{lge99}Lockett, Gauthier, \& Elitzur 1999).  This means the
age of the remnant must fall between $(2-10)\times10^{4}$~y.  These
values are within an order of magnitude of the estimated age of
magnetars, $\sim10,000$~y (\markcite{thomp96}Thompson \& Duncan 1996),
and they imply a projected velocity for the magnetar between 40 and
200~km/s.  These are reasonable values for the projected velocity of a
neutron star.

A distance of 11~kpc also does not demand an excessive energy budget
for these bursts.  Bursts from SGR~1806--20 have been observed to
reach total peak luminosities of $\sim10^{42}$ erg/s, with a third of
the emission above 30~keV (Fenimore et al. 1994).  The two bursts
detected by the ASM from SGR~1627--41, assuming isotropic emission
from an 11~kpc distance, reach peak 5--12~keV luminosities of 2 and
3~$\times10^{39}$~erg/s, only 0.1\% of the above total.

\acknowledgments

We thank Kevin Hurley for providing the IPN annulus and Peter Woods
for providing the {\it BeppoSAX} source location.  DS wishes to thank
Derek Fox, Bryan Gaensler and Vicky Kaspi for helpful discussions.
Support for this work was provided in part by NASA Contract
NAS5-30612.  The MOST is operated by the University of Sydney with
support from the Australian Research Council and the Science
Foundation for Physics within the University of Sydney.

\end{document}